\documentclass[utf8]{frontiersSCNS} 

\usepackage{url,hyperref,lineno,microtype,subcaption}
\usepackage[onehalfspacing]{setspace}
\usepackage{graphicx}
\usepackage{xcolor}
\usepackage{natbib}
\usepackage{txfonts}
\usepackage{color} 
\usepackage{multirow}
\usepackage{rotating}
\usepackage{lscape}
\usepackage{enumerate}
\usepackage{hyperref}
\hypersetup{colorlinks=true, linkcolor=red, citecolor = blue, filecolor=cyan,  urlcolor=magenta,}

\chardef\us=`\_

\def\keyFont{\fontsize{8}{11}\helveticabold }
\def\firstAuthorLast{Sheoran {et~al.}} 
\def\Authors{Jyoti Sheoran\,$^{1}$, Vaibhav Pant\,$^{1}$, Ritesh Patel\,$^{1,2}$, and Dipankar Banerjee\,$^{1,3,4}$}
\def\Address{$^{1}$Aryabhatta Research Institute of Observational Sciences, Nainital 263002, India \\
$^{2}$Southwest Research Institute, 1050 Walnut Street, Suite 300, Boulder, CO 80302, USA\\
$^{3}$Indian Institute of Astrophysics, 2nd Block Koramangala, Bangalore 560034, India \\
$^{4}$Center of Excellence in Space Science, IISER Kolkata, Kolkata 741246, India}
\def\corrAuthor{Jyoti Sheoran, Vaibhav Pant}

\def\corrEmail{jyotisheoran@aries.res.in, vaibhav.pant@aries.res.in}

\begin{document}
\onecolumn
\firstpage{1}

\title[CME Thermodynamics in the Inner Corona]{Evolution of the Thermodynamic Properties of a Coronal Mass Ejection in the Inner Corona} 

\author[\firstAuthorLast ]{\Authors} 
\address{} 
\correspondance{} 

\extraAuth{}

\maketitle

\begin{abstract}

\section{}
The thermodynamic evolution of Coronal Mass Ejections (CMEs) in the inner corona ($\leq$ 1.5 R$_{sun} $) is not yet completely understood. In this work, we study the evolution of thermodynamic properties of a CME core observed in the inner corona on July 20, 2017, by combining the MLSO/K-Cor white-light and the MLSO/CoMP Fe XIII 10747 {\AA} line spectroscopic data. We also estimate the emission measure weighted temperature (T$_{EM}$) of the CME core by applying the Differential Emission Measure (DEM) inversion technique on the SDO/AIA six EUV channels data and compare it with the effective temperature (T$_{eff}$) obtained using Fe XIII line width measurements. We find that the T$_{eff}$ and T$_{EM}$ of the CME core show similar variation and remain almost constant as the CME propagates from $\sim$1.05 to 1.35 R$_{sun}$. The temperature of the CME core is of the order of million-degree kelvin, indicating that it is not associated with a prominence. Further, we estimate the electron density of this CME core using K-Cor polarized brightness (pB) data and found it decreasing by a factor of $\sim$3.6 as the core evolves. An interesting finding is that the temperature of the CME core remains almost constant despite expected adiabatic cooling due to the expansion of the CME core, which suggests that the CME core plasma must be heated as it propagates. We conclude that the expansion of this CME core behaves more like an isothermal than an adiabatic process. 
\tiny
 \keyFont{ \section{Keywords:} Solar Atmosphere, Corona, Coronal Mass Ejections (CMEs), Spectroscopy, Thermodynamics} 
\end{abstract}

\section{Introduction} 
\label{sec:intr}

Coronal mass ejections (CMEs) are large structures of plasma and magnetic fields ejected from the solar atmosphere into the heliosphere (\citealt{1984Hundhausen}; \citealt{2012Webb}). CMEs are the major drivers of space weather because they can cause interplanetary disturbances and shock waves that can lead to the disruption of a range of technologies on Earth (\citealt{1993Gosling}; \citealt{2006Schwenn}; \citealt{2007Pulkkinen}). Thus, it is important to understand their evolution in the solar atmosphere. CMEs have been studied for several decades using remote sensing and in-situ instruments. To study the evolution of CMEs, one uses data from space and ground-based instruments such as the Sun Earth Connection Coronal and Heliospheric Investigation (SECCHI; \citealt{Howard2008}), COR-1, COR-2 and Extreme ultraviolet imager (EUVI; \citealt{2004Wuelser}) instruments on board Solar Terrestrial Relations Observatory (STEREO; \citealt{Kaiser2005}; \citealt{2012Vourlidas}), the Large Angle Spectrscopic COronagraph (LASCO; \citealt{1992Bruekner}) coronagraphs and Extreme-Ultraviolet Imaging Telescope (EIT; \citealt{1995D}) on board the Solar and Heliospheric Observatory (SOHO; \citealt{1995Domingo}), the Mark IV, K-Coronagraph (K-Cor), Coronal Multi-channel Polarimeter (CoMP; \citealt{Tomczyk2008}) instruments at the Mauna Loa Solar Observatory (MLSO; \citealt{2003Elmore}), Atmospheric Imaging Assembly (AIA; \citealt{Lemen2012}) telescopes onboard the Solar Dynamic Observatory (SDO) and the EUV Imaging Spectrometer (EIS; \citealt{2007Culhane}) onboard Hinode. Most of earlier studies focused on the signature of origins of CMEs, determination of their mass/density, dynamic evolution, and the connection between magnetic flux ropes measured in-situ and CME's morphology observed by the white-light coronagraphs and heliospheric imagers. Nevertheless, from white-light data, one cannot estimate the plasma properties of CMEs, such as plasma temperature and elemental composition. 
 CMEs may contain a cool ($\sim 10^4$ K) chromospheric material (prominence/filament), a hot coronal material ($\sim 10^6$ K), and flare plasma ($\sim 10^7$ K), thus, giving rise to the emission in the broad wavelength range. Many processes occurring during CME propagation can interchange different forms of energies (electromagnetic, kinetic, potential, thermal, etc.), causing plasma heating or cooling (\citealt{2007Bemporad}). Thus, to understand these processes, it is crucial to study the evolution of thermodynamic properties (such as density, temperature, thermal pressure, etc.) of CMEs, as it propagates. 
 
However, the evolution of CMEs in the inner corona ($\leq$ 1.5 R$_{sun} $) is not yet completely understood, primarily due to the lack of continuous observations of the inner corona. The dynamics of CMEs in the inner corona are important because this region exhibits kinematics of CME, such as rapid expansion, impulsive acceleration, etc. (\citealt{2001Zhang}, \citeyear{2004Zhang}; \citealt{2003Gallagher}; \citealt{2010Temmer}; \citealt{2011Ben}; \citealt{2011Joshi}; \citealt{2019Sarkar}; \citealt{2020Cremades}; \citealt{2020Majumdar}, \citeyear{2021Majumdar}, \citeyear{2022Majumdar}). However, simultaneous spectroscopy and white-light imaging in the inner corona are needed to improve our understanding of the temperature and kinematics evolution of the CMEs during their initial phase of propagation. There are few studies using ultraviolet (UV) and extreme UV (EUV) imaging and spectral observations, X-ray, and in situ observations where the thermodynamic properties of the CMEs have been studied.
 
The spectroscopic UV observations of CMEs acquired by the Ultraviolet Coronagraph Spectrometer (UVCS; \citealt{1995Kohl}) on board the SOHO has helped us to understand the evolution of CME plasma physical parameters such as densities, temperatures, abundances in the CME core and leading front, and their 3-D velocity structures in the corona from 1.5 to 10 R$_{sun}$. In this heliocentric distance range, the density of the CME leading front (LF) is found to be in the range 10$^4$ - 10$^6$ cm$^{-3}$ (\citealt{2003Ciaravella}, \citeyear{2005Ciaravella}), and the temperature has been inferred ranging from 6.0 $\times$ 10$^3$ to 2.0 $\times$ 10$^6$ K at 1.5 R$_{sun}$ (\citealt{1997Ciaravella}; \citealt{2007Bemporad}). The study by \citet{2022Bemporad} found CME LF density of order 1 $\times$ 10$^7$ cm$^{-3}$ and peak temperature  $\simeq$ 1.9 $\times$ 10$^6$ K at 1.6 R$_{sun}$, and density $\simeq$ 7 $\times$ 10$^6$ cm$^{-3}$ and peak temperature  $\simeq$ 2.1 $\times$ 10$^6$ K at 1.9 R$_{sun}$. The bright cores of CMEs are usually believed to contain the filament/prominence material. The densities of the CME core have been found to range from 1.4 $\times$ 10$^6$ to 7.0 $\times$ 10$^8$ cm$^{-3}$ at 1.3 R$_{sun}$, and density decreases from 1.3 $\times$ 10$^6$ to 4.0 $\times$ 10$^7$ cm$^{-3}$ at 3.0 R$_{sun}$ (\citealt{2001Akmal}; \citealt{2004Raymond}). However, \citet{2022Bemporad} found CME core density of order 1.1 $\times$ 10$^7$ cm$^{-3}$ at 1.6 R$_{sun}$ and 9 $\times$ 10$^6$ cm$^{-3}$ at 1.9 R$_{sun}$, and core peak temperature  $\simeq$ 2.4 $\times$ 10$^6$ K at 1.6 R$_{sun}$ and $\simeq$ 3.2 $\times$ 10$^6$ K at 1.9 R$_{sun}$. Furthermore, these studies were limited to a certain heliocentric distance and/or a particular time.
 
Furthermore, the differential emission measure (DEM) analysis has also been applied to diagnose the physical properties of CMEs. Many CME eruptions have been studied using UV-EUV imagers such as the EIT on board SOHO, the STEREO/EUVI instruments, the AIA telescopes on board SDO, and Hinode/EIS spectroscopic observations. \citealt{2018Long} studied the evolution of a coronal cavity using spectroscopic Hinode/EIS and broadband SDO/AIA observations. The DEM inversion techniques on EUV images have enabled us to infer the 2-D distribution of plasma temperatures and densities inside CMEs. The main results emerge from these studies are that the CME core temperature ranges from 0.8 - 2.0 MK for CMEs associated with prominence eruption, and the core temperature was found to be $\geq $8.0 MK for CMEs associated with a flux rope (\citealt{2016Chmiel}). It was found that the CME core regions were heated during the CME eruption (\citealt{2012Cheng}; \citealt{2010Landi}; \citealt{2013Hannah}), presumably through magnetic reconnection.

In most CME thermodynamic models, the evolution of thermal energy is derived under the assumption of adiabatic expansion of plasma material (\citealt{2017Durand}). However, many studies addressing the energy budget of CME plasma demonstrate the presence of an additional energy source responsible for the heating of the CME plasma. The additional heating source provides thermal energy that was found to be comparable to the total kinetic and potential energies gained by the eruptions (\citealt{2001Akmal}; \citealt{2003Ciaravella}; \citealt{2011Murphy}). While in some cases, it was found to be much larger than the total kinetic and potential energies gained by the eruptions (\citealt{2009Lee}; \citealt{2010Landi}). Also, based on in-situ observations of ICMEs, the polytropic index ($\gamma$) of ICME plasma was suggested to be of the order of 1.1 to 1.3 from 0.3 and 20 AU (\citealt{2005Liu}, \citeyear{2006Liu}), implying the local heating of ICMEs plasma. \citealt{2018Mishra} found that the polytropic index of a CME plasma decreased from 1.8 to 1.35 as the CME moved from 5.9 to 13.9 R$_{sun}$, implying that CME first released heat to reach an adiabatic state and then absorbed heat. However, our understanding of the evolution of thermodynamic properties of CME during its propagation is still limited. 

Despite this, we still lack an understanding of how the thermodynamic properties of CMEs change during their propagation in the inner corona (up to $\sim$ 1.5 R$_{sun} $). Using UVCS spectroscopic data, the physical parameters of CME plasma have been obtained at fixed heliocentric distances starting from 1.5  to 10 R$_{sun}$. However, the DEM analysis using SOHO/EIT, STEREO/EUVI, $\&$ SDO/AIA, has enabled us to estimate the physical parameters of CME in the inner corona (up to $\sim$ 1.5 R$_{sun} $). Nevertheless, in none of these studies, the continuous evolution of the thermodynamic parameters of CME structures in the inner corona has been studied. Hence, the motivation behind the work presented here. In this paper, by combining the K-Cor white-light data and the CoMP Fe XIII 10747 {\AA} line spectroscopic data, we present the spectroscopic diagnostics of the temperature and density of a CME core observed in the inner corona on July 20, 2017. We studied the continuous evolution of the thermodynamic properties, such as the temperature and density of the CME core from $\sim$ 1.05 to 1.35 R$_{sun}$. In Section \ref{sec:Obs}, we provide the details of the observations. In Section \ref{sec:analysis}, we describe the methods used to derive the temperature and density of this CME core and present the main results, followed by a summary \& discussion in Section \ref{sec:sd}.
\section{Observations} \label{sec:Obs}
 \begin{figure*}[!ht]
    \includegraphics[scale = 0.592]{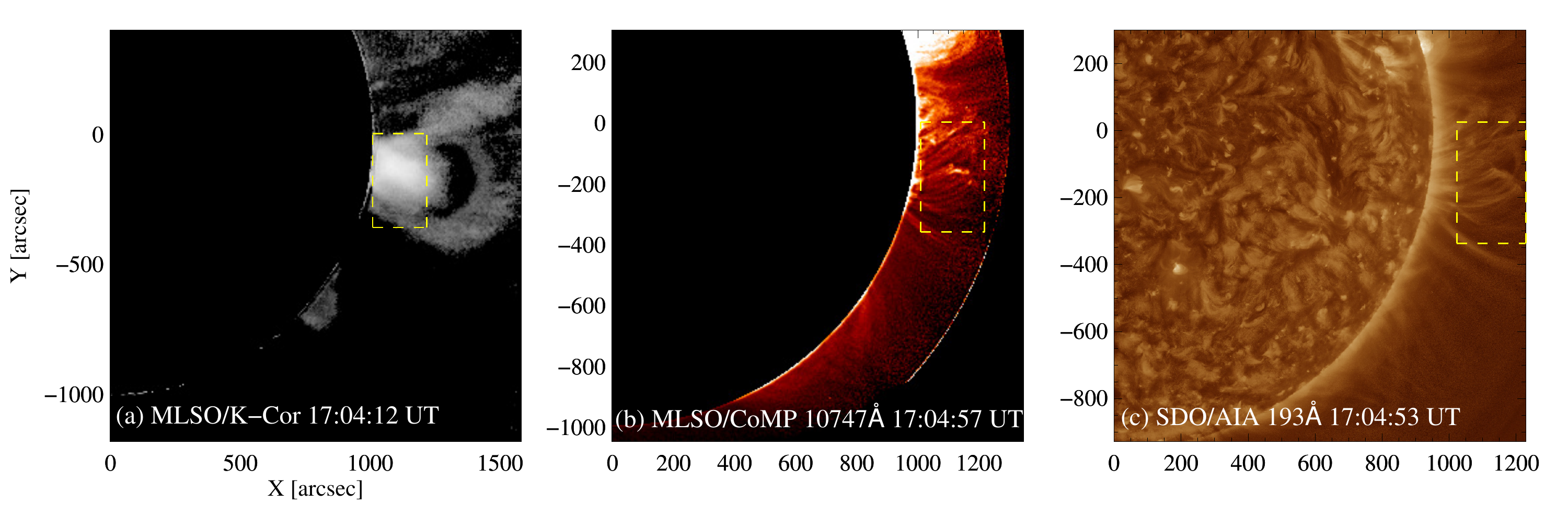}
    \caption{(a) July 20, 2017, CME as observed by the K-Cor. All three parts of the CME are clearly visible. The yellow rectangular box shows the ROI chosen for the analysis. (b) The CoMP 10747 {\AA} channel enhanced intensity image. When CoMP observation started, only the core of the CME was visible in the CoMP FOV. The yellow box is the CoMP ROI aligned with the K-Cor ROI shown in Figure (a). (c) The AIA 193 {\AA} channel image processed using MGN to enhance the CME core structures. 
    The yellow box is the AIA ROI aligned with the CoMP ROI shown in Figure (b). (An animation of the evolution of the CME core in the K-Cor, CoMP 10747 {\AA}, and AIA 193 {\AA} channel is available in the Electronic Supplementary Material. The yellow circle in the animation represents the CoMP FOV.)}
    \label{fig:kca}
\end{figure*}
On July 20, 2017,  the MLSO/K-Cor observed a CME propagating outward from the west limb starting at around 17:00 UT. 
The event is observed with multiple instruments, including the K-Cor and CoMP, the AIA onboard SDO, and the white-light coronagraph (COR-1A) of SECCHI onboard STEREO. The K-Cor records polarized brightness (pB) images of the inner corona from 1.05 - 3 R$_{sun}$ in 720 - 750 nm passband with a cadence of 15 seconds. The K-Cor images have a pixel scale of $\approx5.6$ arcsecs pixel$^{-1}$. We used white-light level-2 pB data to estimate the electron density. Figure \ref{fig:kca}(a) shows the three-part CME in the K-Cor FOV.

The eruption was also seen in the CoMP Fe XIII 10747 {\AA} channel. The CoMP takes observation of the polarization state at a few spectral locations across the profiles of three infrared lines (Fe XIII 10747 {\AA},  10798 {\AA} \& He I 10830 {\AA}) using a narrow-band tunable filter. It has a field of view (FOV) of 1.05 - 1.4 R$_{sun}$, a pixel-scale of 4.46 arcsecs pixel$^{-1} $, and a typical cadence of 30 seconds. The details of the acquisition and reduction of CoMP data are described in  \citealt{Tomczyk2008}, and the calculation process is given by \citealt{Tian_2013}. We used Fe XIII 10747 {\AA} level 2 dynamic data, which contains line peak intensity, edge enhanced line peak intensity, line-of-sight (LOS) Doppler velocity, and line width images. On this day, the CoMP observation started at around 16:57 UT, and only the core of this CME was visible in the CoMP FOV, as shown in Figure \ref{fig:kca}(b). This eruption was seen as a braided structure moving out in CoMP Fe XIII 10747 {\AA} channel and caused a significant enhancement in Fe XIII line width.

This eruption was also recorded in all SDO/AIA EUV channels from around 15:00 to 18:00 UT. The AIA takes full-disk images of the corona and transition region up to 1.3 R$_{sun}$ simultaneously in seven extreme-ultraviolet (EUV) narrow-band filters. AIA has a pixel-scale of 0.6 arcsecs pixel$^{-1}$ and a cadence of 12 seconds. Using \textrm{aia\_prep.pro} AIA images were normalized by their exposure times, rotated, and re-scaled to ensure that each pixel is examining the same spatial location across the different wavelengths. We also checked AIA 304 {\AA} data-set to determine if this CME has an associated filament. No filament could be seen in AIA 304 {\AA} channel prior to the period of study. We used 94, 131, 171, 193, 211, and 335 {\AA} pass bands images, sensitive over a temperature range from 10$^5$ K to 10$^7$ K for the DEM analysis. Since AIA 193 {\AA} and CoMP 10747 {\AA} images have similar temperature responses, the core of this CME as seen in AIA 193 {\AA} channel, is shown in Figure \ref{fig:kca}(c). The
AIA 193 {\AA} channel image has been processed using Multiscale Gaussian Normalization \citep[MGN:][]{2014Morgan} to enhance the CME core structures.

The yellow rectangular box in all panels of Figure \ref{fig:kca} corresponds to the region of interest (ROI) chosen for the analysis. We have aligned CoMP 10747 {\AA} ROI with respect to AIA 193 {\AA} ROI. 
The yellow dashed lines in Figure \ref{fig:dem_map} (a) 
indicate the position of four artificial slits at the same location in both AIA and CoMP ROIs. Each slit is approximately ten arcsec wide and the location of these slits is such that they cover almost the entire CME core. Since K-Cor images capture white-light and CoMP images capture narrow-band emission, we could not use intensity/brightness to align their FOVs as the features look different in white-light and narrow-band emission images. The CoMP and the K-Cor both have 1.05 R$_{sun}$ occulter. We use it as a reference to align the CoMP \& the K-Cor ROIs.

This CME is also observed by the STEREO/COR-1A coronagraph. The COR-1A coronagraph offers high-cadence (5 minutes) data with FOV from 1.4-4 R$_{sun}$, and a pixel scale of 7.5 arcsecs pixel$^{-1}$. There is a filament behind the limb, which could be seen in STEREO EUVI 304 {\AA} channel around 11:46 UT on July 20, 2017 which erupts around 12:46 UT, seen in COR-1A FOV  $\sim$13:00 UT. Around 15:00 UT, a braided structure is seen to rise in EUVI 304 Å and 195 Å FOV, which propagates out and a CME appears (case event for this study) in COR-1 FOV at around 16:40 UT. Therefore, our case event does not have a filament associated with it. To determine the parameters like LOS depth and volume of this CME, we used the graduated cylindrical shell (GCS) model (\citealt{2009Thernisien}) to fit this CME using the K-Cor \& COR-1A vantage point observations. For GCS fitting, we have used K-Cor white light level 2 Normalizing Radially Graded Filter (NGRF; \citealt{2006morgan}) data and COR-1 level 1 data filtered using Simple Radial Gradient Filter (SiRGraF; \citealt{2022ritesh}).
 
\section{Analysis and Results } 
\label{sec:analysis}

\subsection{Determination of the Electron Temperature } 
\label{subsec:Temp}

In this work, we determined the two-dimensional plasma temperature distribution across the CME core using two methods. First, using DEM analysis on AIA six EUV channels data, and second, using the line broadening of Fe XIII emission line centered at 10747 {\AA}.
\subsubsection{Temperature using the Differential Emission Measure Analysis} \label{subsec:DEM_analysis}

To infer the DEM from AIA six EUV channels data, we applied the inversion technique developed by \citealt{2015Cheung}. For the inversion, we used a temperature grid spanning log T/K $\in [5.7, 7.7]$ with a grid spacing of log T = 0.1. The emission measure in each temperature grid was obtained using \textrm{aia\_sparse\_em\_init.pro} in the SolarSoftware (SSWIDL) package. The total emission measure (EM) was then obtained by summing the EM in all temperature bins:
 
\begin{equation} 
    \label{eq3}
    EM = \sum_{j}^{n}EM_{j},
\end{equation}
    
where EM$_j$ is the EM contained in the jth temperature bin, and n is the total number of bins (in our case n = 21). 
 The mean temperature can be estimated by the emission-weighted temperature, 
\begin{equation} 
    \label{eq4}
    log\:T_{EM} = EM^{-1}\left[\sum_{j}^{n}EM_{j}\:log\:T_{j}\right],
    \end{equation}

\begin{equation} 
    \label{eq5}
    W_{EM}^2 = EM^{-1}\left[\sum_{j}^{n}EM_{j}\bigl(log\:T_{j}\:-\:log\:T_{EM}\bigr)^2\right],
\end{equation}

here log T$_{EM}$ is the EM-weighted log temperature and W$_{EM}$ is the effective width of the distribution in log T space.

\begin{figure*}[ht!]
    \centering
    \includegraphics[width=0.95\textwidth]{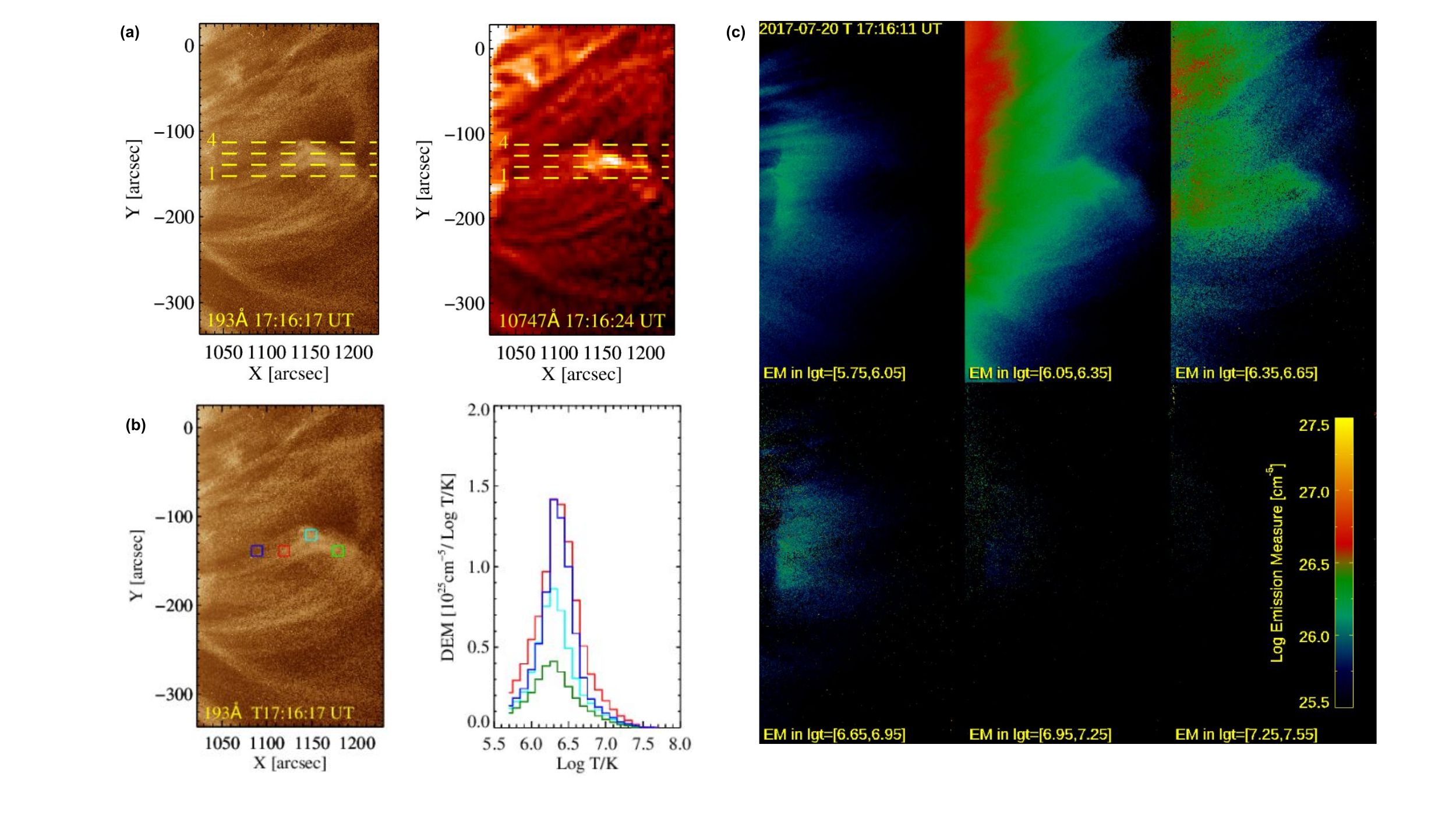}
    \caption{(a) The left panel shows the AIA ROI (shown by yellow box in Figure \ref{fig:kca} (c)), and right panel shows the CoMP ROI (shown by yellow box in Figure \ref{fig:kca} (b)). The yellow dashed lines in both panels show the locations of four co-spatial slits chosen in the two ROIs. (b) The left panel shows AIA 193 {\AA} ROI image at 2017-07-20 T 17:16:17 UT. The colored boxes show the selected sub-regions used to reconstruct the DEM curves. The DEM curves for these box regions are shown in the right panel. The DEM in all regions peaked at around log T/K $\sim$ 6.3. 
    (c) EM maps of the AIA ROI at 2017-07-20 T 17:16:11 UT. 
    The color coding indicates the total EM contained within a log temperature range indicated in the bottom left corner of each panel. (An animation is available in the Electronic Supplementary Material.)
    }
    \label{fig:dem_map}
\end{figure*}

\begin{figure*}[!ht]
    \centering
    \includegraphics[angle=0,scale=1.68]{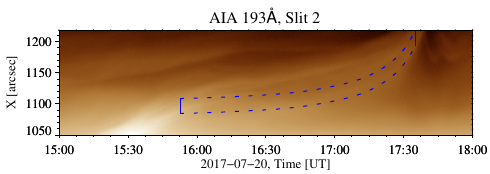}
    \includegraphics[angle=0,scale=1.68]{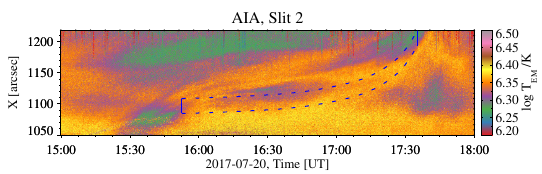}
    \includegraphics[angle=0,scale=1.68]{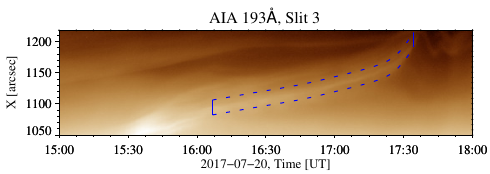}
    \includegraphics[angle=0,scale=1.68]{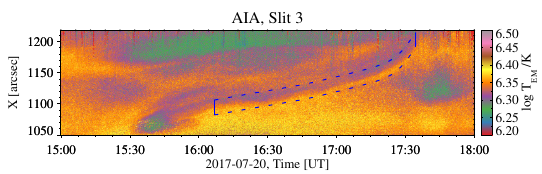}
    \caption{The left panel shows AIA 193 {\AA} intensity space-time maps, and the right panel shows the EM-weighted log temperature space-time maps for all slits. We visually traced the eruption along each slit intensity space-time map shown by the blue dashed curve in the left panel, which is fitted using the cubic spline interpolation method. The same curve is overplotted over log T$_{EM}$ space-time map of the respective slit.}
    \label{fig:ilgt_ht_var}
\end{figure*}

We created DEM maps at each pixel for the logarithmic temperature spanning from 5.7 to 7.7. 
The colored boxes in the left panel of  Figure \ref{fig:dem_map} (b) show the selected sub-regions of AIA ROI used to reconstruct the DEM curves. Note that we have processed AIA 193 {\AA} channel images using the MGN technique to enhance the CME core structures, the MGN technique has not been used for DEM analysis.
 The DEM curves for these box regions are shown in the right panel. The DEM in all regions peaked at around log T/K $\sim$6.3. Figure \ref{fig:dem_map} (c) show the EM maps of the AIA ROI at 2017-07-20 T 17:16:11 UT. We can see that the EM of the core of the CME is mostly confined in log T/K $\in$ [6.05, 6.65]. Then using Equation (\ref{eq4}) we created EM-weighted log temperature (log T$_{EM}$) maps of the AIA ROI. We obtained the intensity and log T$_{EM}$ values along the AIA slits (shown in the right panel of Figure \ref{fig:dem_map} (a)) and took their averaged over the width of each slit to ensure a good signal-to-noise ratio. To illustrate the full evolution of the core of the CME, we constructed space-time maps for all four slits. The left panel of Figure \ref{fig:ilgt_ht_var} shows the AIA 193 {\AA} intensity space-time maps, and the right panel shows the log T$_{EM}$ space-time maps for slit 2 and 3. We visually inspected the AIA 193 {\AA} intensity space-time map of slit 2 and tracked the eruption along this slit as shown by the blue dashed curve in the top left panel of Figure \ref{fig:ilgt_ht_var}, which we fitted using the cubic spline interpolation method. Using these height-time values, we fitted the spline curve on log T$_{EM}$ space-time map of slit 2 as shown in the top right panel of Figure \ref{fig:ilgt_ht_var} by blue dashed curve and obtained the values of log T$_{EM}$ and W$_{EM}$  (W$_{EM}$ is calculated using Equation (\ref{eq5})) along this curve. The same procedure is repeated for the remaining slits. Note that the spline fitted curve has approx 24 arcsec spatial extent at a particular time. 

\begin{figure*}[ht!]
    \centering
    \includegraphics[width=1\textwidth]{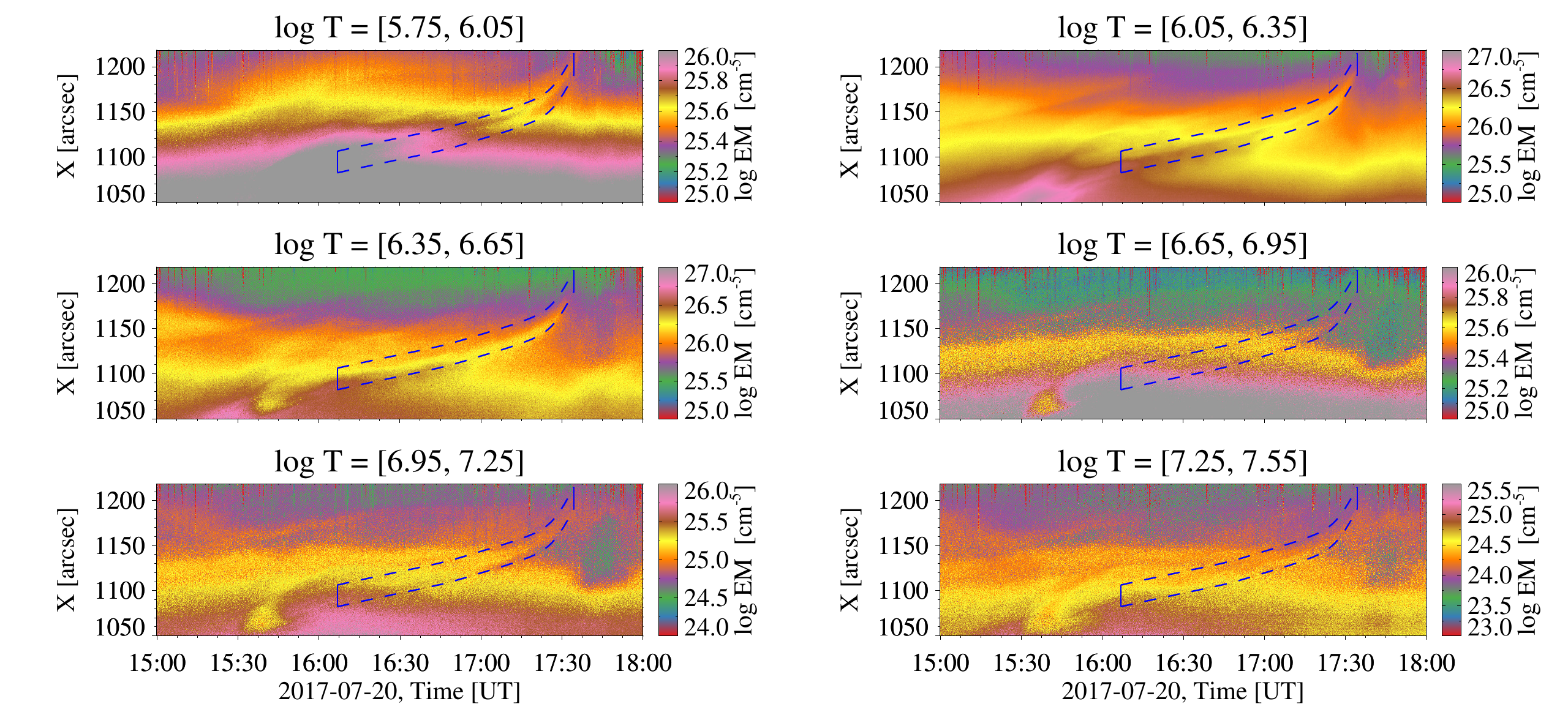}
    \caption{ The temporal variation in  log EM  along slit 3 for different temperature bins. The blue dashed curve in all panels of is the same spline curve as in the bottom panel of Figure \ref{fig:ilgt_ht_var}.
    }
    \label{fig:em_xt}
\end{figure*}

The calculation of EM-weighted temperature is based on the assumption that the plasma is isothermal along the LOS. However, the left panel of Figure \ref{fig:dem_map} (b) demonstrates that the emission measure has a spread over a range of temperatures. To investigate the temporal evolution of the CME core emission in different temperature bins, we produce space-time maps of log EM for slit 3. Figure \ref{fig:em_xt} shows temporal variation in log EM  along slit 3 for different temperature bins. The space-time maps show that the CME core rises gradually, and at $\sim$17:30 UT, the CME core moves out of the AIA FOV. The CME core appears in most of the temperature bins indicating that the CME core is multi-thermal. However, the EM of the CME core is mainly confined in log T/K $\in$ [6.05, 6.65]. The blue dashed curve in all panels of Figure \ref{fig:em_xt} is the same spline curve as in the bottom panel of Figure \ref{fig:ilgt_ht_var}.

\subsubsection{Effective temperature from Fe XIII Line Width}\label{lw_analysis}

\begin{figure*}[!ht]
    \centering
     \includegraphics[scale = 1.6]{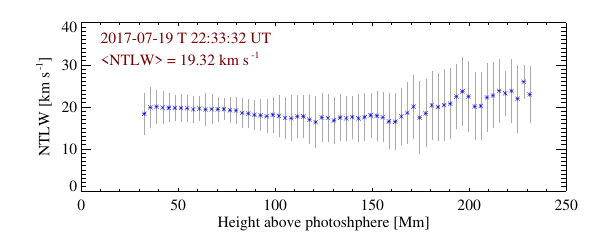}
     \caption{The variation of NTLW with height above the solar surface. }
     \label{fig:ntw_h}
 \end{figure*}

Using the Fe XIII line width and considering that the broadening of the line may occur due to the thermal motions of ions, the non-thermal motions in the corona, the instrumental broadening, the expansion of CME as it moves outward, and the additional turbulence created by the CME propagation in the corona, the CME plasma temperature can be obtained. In order to calculate the non-thermal line width (NTLW) in the corona, we used the previous day, i.e., July 19, 2017, CoMP Fe XIII 10747 {\AA} line width data when CME is not present. We subtract the thermal width (21 km s$^{-1}$, corresponding to the peak formation temperature of Fe XIII of $\sim$1.6 MK) and instrumental width (21 km s$^{-1}$) calculated by \citealt{2015Morton} from the line width obtained for each pixel (\citealt{2012McIntosh}). To obtain the radial profile of NTLW, we took a median value of NTLW of pixels lying along 40 degrees (to ensure a good signal-to-noise ratio) about the equator at each radius. Figure \ref{fig:ntw_h} shows the variation of the NTLW with height above the solar surface. We found that the NTLW did not vary significantly with height. Therefore, for our calculations, we used the mean value of NTLW, 19.32 km s$^{-1}$. 
We subtract non-thermal width (NTLW) and instrument width in quadrature from the total line width; the residual width is Doppler/thermal width (i.e., width due to thermal motions of ions). Furthermore, the plasma temperature can be estimated using the following Equation:

\begin{equation} 
    \label{eq6}
    T = \frac{1}{2}\frac{m}{k_B}v_{1/e}^2,
\end{equation}
    
where m is the mass of the Fe XIII ion, k$_{B}$ is the Boltzmann constant, and v$_{1/e}$ is the velocity derived from the Doppler half-width, $\Delta\lambda_{1/e}$. Note that we did not take into account the broadening of the line due to the expansion of the CME core and the additional turbulence created by the propagation of the CME. Hence Equation \ref{eq6} provides the upper limit to the plasma temperature. We refer to it as effective temperature, T$_{eff}$.

\begin{figure*}[ht!]
    \centering
     \includegraphics[angle=0,scale=1.75]{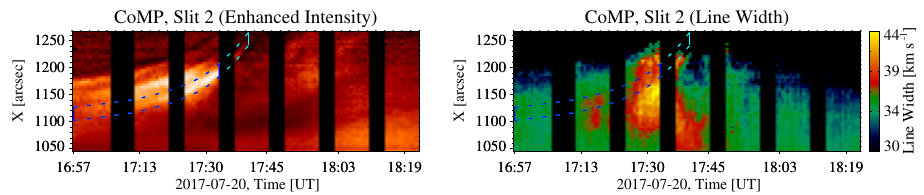}
     \includegraphics[angle=0,scale=1.75]{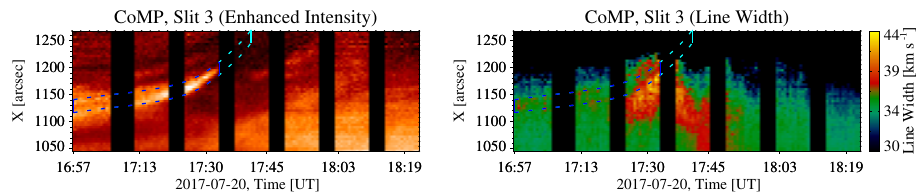}
     \caption{The left panel shows the CoMP Fe XIII 10747 {\AA} enhanced-intensity space-time maps, and the right panel shows the Fe XIII line width space-time maps. The blue dashed curve in each slit space-time map is fitted using the height-time values obtained from spline fitting AIA 193 {\AA} intensity space-time map of the respective slit. The eruption is further tracked in the CoMP FOV outside the AIA FOV and is shown by the cyan curve. Black stripes represent gaps in data.}
    \label{fig:comp_slits_map}
\end{figure*}

We selected an ROI from the CoMP (Figure \ref{fig:kca} (b)) image same as AIA 193 ROI and placed four artificial slits co-spatial with AIA slits (Figure \ref{fig:dem_map} (a)). We averaged the Fe XIII line enhanced intensity and line width over the width of the slits and created the space-time maps of line enhanced intensity and total width as shown in the left and right panels of Figure \ref{fig:comp_slits_map}, respectively. Then, using the height-time values along the spline fitted curve on AIA 193 slit 2 intensity space-time map, we fitted the Fe XIII enhanced intensity and line width space-time maps of slit 2, which is shown by the blue dashed curve in the top panel of Figure \ref{fig:comp_slits_map}. Since the CoMP has a larger FOV than AIA, we visually tracked the eruption in CoMP FOV outside the AIA FOV shown by the cyan curve in Figure  \ref{fig:comp_slits_map}. The line width values were obtained along the blue and the cyan curves. Moreover, to calculate the Doppler/thermal width values, the NTLW (19 km s$^{-1}$) and instrumental width (21 km s$^{-1}$) were subtracted from the total line width values. The effective temperature was calculated using Equation (\ref{eq6}). This procedure is repeated for the remaining slits also.

\begin{figure*}[ht!]
    \centering
    
    \includegraphics[angle=0,scale=0.5]{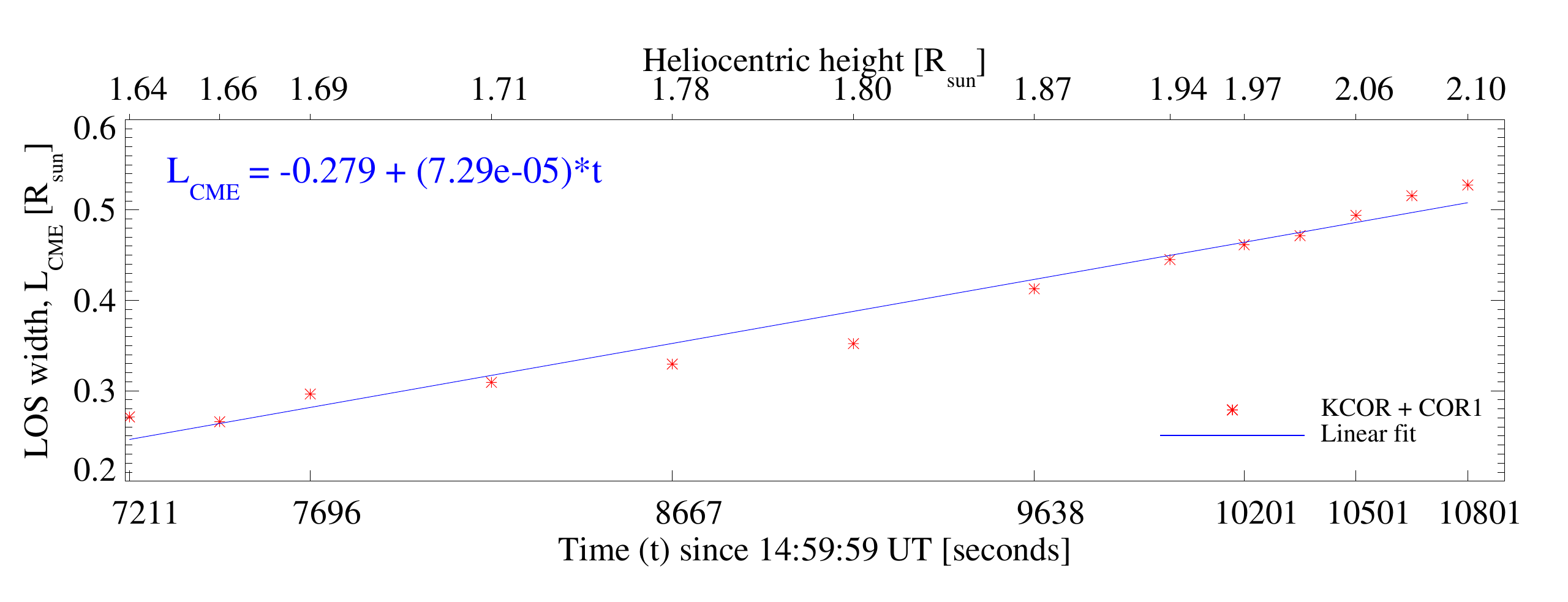}

    \caption{ The variation of LOS width (L$_{CME}$) of the CME core with time or heliocentric height.}
    \label{fig:lcme}
    \end{figure*}

In addition to this, expansion of the CME core and additional turbulence created by the CME propagation will also contribute to the non-thermal broadening.
To model the expansion of the CME core, we used the graduated cylindrical shell (GCS) model developed by \citealt{2009Thernisien}; \citealt {2011Thernisien} and followed the fitting procedure described in \citealt{2022Majumdar}. We used the GCS ice-cream cone model to fit the core of the CME. Figure \ref{fig:lcme} shows the variation of the LOS nominal depth of the CME core plasma (L$_{CME}$) with time and heliocentric distance of the CME core front. We found that the L$_{CME}$ of the CME core increases almost two times as the core evolves during this period. The expansion velocity of the CME core is found to be approx. 50.76 km/s, which is larger than the total line width. Since the features look different in the white-light $\&$ IR emission band ( Fe XIII 10747 {\AA}), we conclude that we may not be modelling the same feature using the GCS model fit, which we are tracking in IR emission. Therefore, the expansion factor for these two wavelength regimes would be different, which makes it difficult to apply the expansion factor obtained from the GCS model to account for the broadening of the Fe XIII 10747 {\AA} line due to the expansion of the CME core.

\begin{figure}[ht!]
    \centering
  
    \includegraphics[angle=0,scale=1.77]{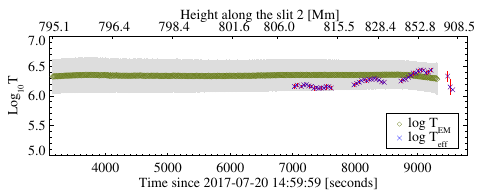}
    \includegraphics[angle=0,scale=1.77]{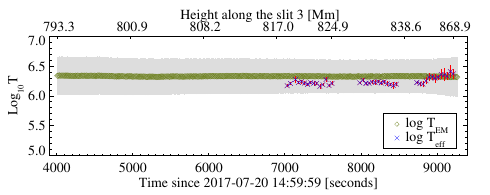}

    \caption{The evolution of log temperature of the CME core with height and time. The green diamond symbols show the variation of the EM-weighted log temperature (log T$_{EM}$) averaged over the spatial direction along the blue curve shown in Figure \ref{fig:ilgt_ht_var}. The grey shaded region is the effective width of the distribution in log T space. The blue cross symbols show the variation of log effective temperature averaged over the spatial direction along the blue and the cyan curves shown in Figure \ref{fig:comp_slits_map}, and the red bars are one sigma deviation in log T$_{eff}$ values.}
    \label{fig:T_ht_var}
\end{figure}

Figure \ref{fig:T_ht_var} shows the evolution of log temperature of the CME core with height and time. The green diamond symbols show the variation of the EM-weighted log temperature (log T$_{EM}$) averaged over the spatial direction along the blue curve shown in Figure \ref{fig:ilgt_ht_var}. The shaded grey region shows the effective width of the distribution in log T space, which is obtained by adding W$_{EM}$ and the standard deviation for this spline-fitted curve. We found that within this region of uncertainty, the EM-weighted temperature of the core of the CME remains almost constant as CME evolves, and the mean value of log T$_{EM}$/K of the core of the CME is found to be in the range of 6.28 - 6.36. The blue cross symbols in the Figure \ref{fig:T_ht_var} show the variation of log effective temperature averaged over the spatial direction along the blue and the cyan curves shown in Figure \ref{fig:comp_slits_map}, and the red bars are one sigma deviation in log T$_{eff}$ values. The log effective temperature of the core of the CME has a similar variation to the EM-weighted log temperature and has mean values in the range of 5.97 - 6.57.

\subsection{Estimation of the Electron Density}

  The electron density can be derived using the CoMP Fe XIII 10747 $\&$ 10798 {\AA} density-sensitive line pair. Since there were no good signal-to-noise ratio data frames available during the period of CME under consideration, we could not use this line pair for density calculation. Therefore, we use K-Cor polarized brightness data to derive the electron density. 
 
The polarized brightness observed by the white-light coronagraphs primarily depends on the column electron density (LOS integration of the electron density). Transient phenomena, such as CMEs, cause the intensity (hence, density) enhancements in a sequence of coronagraph images. A suitable pre-event image is subtracted from the frames containing the CME to calculate the CME density. Since no pre-event frame was available for the CME event on July 20, 2017, we chose a frame on July 21, 2017, at around 02:21:59 UT, much later after the CME had passed the K-Cor FOV, and subtracted this image from the CME frames. As a result, the background F-corona and static K-corona are removed,  leaving us with the brightness changes caused by the CME.

The excess column electron density N$_{e}$ (due to the CME) can be estimated by taking the ratio of the excess observed pB (pB$_{obs}$) to the polarized brightness of a single electron (pB$_{e}$) assumed to lie on the plane of sky (POS) (\citealt{2009Colaninno}). The POS assumption is valid for July 20, 2017, CME as indicated by very small LOS Doppler velocities values in CoMP Fe XIII 10747 {\AA} data. The pB$_{e}$ is computed analytically from the scattering geometry using the equations given in \citealt{1966Billings}. We used \textrm{eltheory.pro} in the SolarSoftware (SSWIDL) package to calculate the pB$_{e}$, and the limb darkening coefficient used in this routine was calculated using the Equation (5) of \citealt{1998Hestroffer} for wavelength = 735 nm (for the K-Cor). The CME electron density n$_e$, can be obtained by dividing the derived column density N$_e$ to the LOS nominal depth of the CME plasma L$_{CME}$, n$_e$ = N$_e$/L$_{CME}$ (\citealt{2022Bemporad}).

\begin{figure*}[ht!]
        \centering
        \includegraphics[angle=0,scale=1.76]{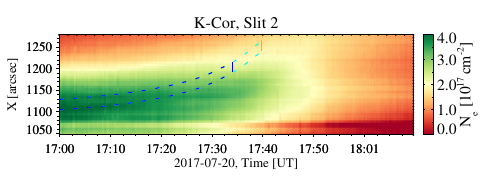}
        \includegraphics[angle=0,scale=1.76]{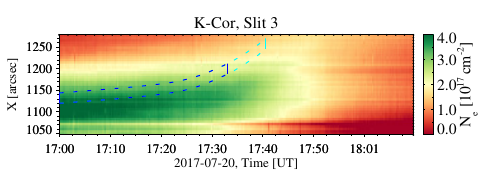}
        \caption{The column electron density space-time maps. The blue and the cyan curves are the same as in Figure \ref{fig:comp_slits_map}.}
        \label{fig:ne_map}
        \end{figure*}

\begin{figure*}[ht!]
    \centering
    \includegraphics[angle=0,scale=1.77]{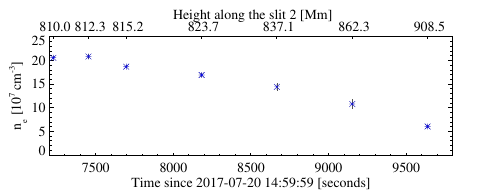}
    \includegraphics[angle=0,scale=1.77]{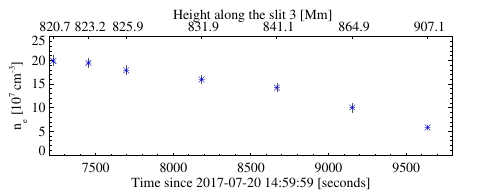}
    \caption{The evolution of electron density (n$_e$) of the CME core with height and time. The blue cross symbols show the variation of n$_e$ averaged over the spatial direction along the blue and the cyan curves shown in Figure \ref{fig:ne_map}, and the dark grey bars are one sigma deviation in n$_e$ values.}
    \label{fig:ne_ht_var}
    \end{figure*}

We determined the column electron density for the entire K-Cor ROI (the yellow box in Figure \ref{fig:kca}(a)) and chose four slits co-spatial with AIA and CoMP slits. The resulting 2-D maps of the column density (N$_e$) for slit 2 $\&$ 3 are shown in Figure \ref{fig:ne_map}. The green region in all panels shows the evolution of the core of the CME with time along the respective slit. The blue and cyan curves are the same as described in section \ref{lw_analysis}. We obtained the values of N$_e$ along the spline fitted curves and divided it by L$_{CME}$  to obtain the CME core electron density values (n$_e$). The variation of the CME core electron density (n$_e$) along the spline fitted curves for slit 2 $\&$ 3 is shown in Figure \ref{fig:ne_ht_var} by blue cross symbols. The dark grey bars show one sigma uncertainty in the electron density. We found that the CME core's density falls by a factor of $\sim$3.6 as the core evolves. The density values of the CME core are in the range (5.85- 20.85)$\times$ 10$^7$ cm$^{-3}$. We also derived the volume of the CME core using the GCS model fitted parameters. The volume of the CME core increases approx. six times while the core evolves in the K-Cor FOV. Hence, the decrease in CME core density is consistent with the volume increase of the CME core. Similar results were also obtained for slits 1 $\&$ 4.

\section{Summary and Discussions} \label{sec:sd}
 In this work, we carried out the spectroscopic diagnostics in the inner corona (1.05 - 1.35 R$_{sun}$) to derive the thermodynamic property of a CME by combining the CoMP Fe XIII 10747 {\AA} line width data and the K-Cor polarized brightness (pB) data. We studied the evolution of a CME core's thermodynamic properties that occurred on July 20, 2017. We obtained the effective temperature of the CME core using the line broadening of the Fe XIII emission line centered at 10747 {\AA}. We have also applied the DEM inversion technique on AIA six EUV channels data to determine the EM - weighted temperature of the CME core and compare it with the effective temperature obtained using Fe XIII line width. The column density of the CME core is derived using the K-Cor pB intensity. To obtain the LOS depth (L$_{CME}$) and volume of the CME core, we used the graduated cylindrical shell (GCS) model to fit this CME core using the two (K-Cor \& COR-1A) vantage point observations. We obtained the electron density of the CME core by dividing the CME core column density by L$_{CME}$ of the CME core plasma. 
 
 We find that within one sigma error, the EM-weighted temperature of the CME core remains almost constant as CME evolves, and the mean log T$_{EM}$/K of the CME core is found to be in the range 6.28 - 6.36. The effective temperature of the CME core also has a similar variation to the EM- weighted temperature, and mean log T$_{eff}$/K has values in the range of 5.97 - 6.57. The non-thermal motions in the corona contribute significantly to the broadening of a spectral line (\citealt{2012McIntosh}; \citealt{2016Brooks}; \citealt{2019Pant}). 
In earlier studies, the plasma temperatures have also been obtained using the width of spectral lines (e.g., H I Ly-$\alpha$ and Ly-$\beta$, C III, etc.) observed in the CMEs (\citealt{2000Ciaravella}; \citealt{2016Heinzel}). Nevertheless, these studies did not take into account the contribution of non-thermal motions in the width of the spectral line. Hence, the plasma temperature limit provided by these studies is not quite accurate. Therefore, it is required to subtract NTLW from the total width of a spectral line (\citealt{2016Heinzel}). In our analysis, we have calculated NTLW and subtracted this from the total width of the Fe XIII 10747 {\AA} line. The consequence is that the CME core's effective and EM-weighted temperatures have similar values within the error bars. Thus, we can conclude that the effective temperature derived from line width by taking into account the non-thermal in the corona and instrumental broadenings is a good measure of CME plasma temperature. The million-degree kelvin temperature of the CME core indicates that the core of this CME is not associated with a prominence material.

We find that the CME core's electron density falls by a factor of $\sim$3.6 as the core evolves and has values in the range (5.85- 20.85)$\times$ 10$^7$ cm$^{-3}$. As pointed out by \citealt{2018Bemporad}, the uncertainty in the density determination using pB is mainly due to the POS assumption (the assumption that all the plasma is located on the POS) and due to the uncertainty in the depth of a CME structure along the LOS (L$_{CME}$). However, the POS assumption is valid for July 20, 2017, CME as indicated by very small LOS Doppler velocities values in CoMP Fe XIII 10747 {\AA} data, and we get a better estimate of L$_{CME}$ by performing GCS model fit to the CME core.

We find that the temperature of the CME core remains almost constant despite expected adiabatic cooling due to the expansion of the CME core, which suggests that the CME core plasma must be heated as it propagates.  
In previous studies based on in-situ observations of ICMEs, the polytropic index ($\gamma$) of ICME plasma is also found to be close to unity implying the isothermal expansion of ICME plasma. $\gamma$ of ICME plasma was suggested to be of the order of 1.1 to 1.3 from 0.3 and 20 au (\citealt{2005Liu}, \citeyear{2006Liu}), indicating local heating of ICME plasma. Furthermore, in few MHD models of CME have also used $\gamma$ close to unity. \citealt{1998Gibson} constructed a theoretical MHD model describing the ejection of a 3-D CME out of the solar corona by making an assumption $\gamma$ $\sim$ 4/3. \citealt{2002Odstrcil} have used $\gamma$ = 1.05 in a coronal 2-D MHD model to simulate the disruption of a sheared helmet streamer launching a CME. \citealt{1996Chen} $\&$ \citealt{2000Krall} have used $\gamma$ = 1.2 in their theoretical treatment describing the initiation, propagation, and driving mechanisms of ICMEs. Thus, we can infer that the expansion of July 20, 2017, CME core behaves more like an isothermal than an adiabatic process during its evolution in the inner corona from 1.05 - 1.35 R$_{sun}$. 
It may also be possible that the thermal force is the internal driver of CME expansion, as highlighted in the study by \citealt{2018Mishra}.
The presence of plasma heating processes occurring during the CME expansion is also reported in many studies in the literature (\citealt{2001Akmal}; \citealt{2003Ciaravella}; \citealt{2009Lee}; \citealt{2010Landi}; \citealt{2011Murphy}; \citealt{2022Bemporad}). The candidate heating mechanisms are briefly discussed in studies by \citealt{1991Kahler, 1996Kumar, 2011Murphy}; however, there is no widely accepted heating mechanism. Moreover, the above conclusion that the CME core is not a filament/prominence and the CME core is continually heated during its early expansion has an important inference on the initiation mechanism of CMEs. Currently, the existing theories of CME initiation fall into two categories: one is based on the ideal MHD instability of magnetic flux rope ( \citealt{2005Torok,2006kliem,2007Fan,2010Aulanier,2014Kliem,2018Amari}), which often contains a filament, and the other is based on magnetic reconnection (\citealt{1999Antiochos,2001Moore,2017Wyper,2021Jiang,2021JiangC}), which does not require a pre-existing flux rope or filament. In the reconnection model, the core of the CME is formed by reconnection, and thus it is continually heated by the reconnection. On the other hand, in the ideal instability-based model, the CME core is the pre-existing flux rope; thus, no heating can be provided during the ideal expansion. Hence, July 20, 2017, CME event supports the reconnection model.

Our work demonstrates the potential of the CoMP and the K-Cor and future multi-channel coronagraphs upgraded-CoMP (UCoMP) at MLSO and Visible Emission Line Coronagraph \citep[VELC:][]{VELC17, IAUS2017} onboard Aditya-L1 
 to study the thermodynamic evolution of CMEs in the inner corona. The UCoMP has the capability to perform simultaneous 2-D imaging and high-resolution spectroscopy in the inner corona. It has a FOV of 1.03 - 1.95 R$_{sun}$ and a spatial resolution of 6 arcsecs (3 arcsecs/pixel). The UCoMP has seven coronal emission lines (FeXIV 530.3 nm, FeX 637.4 nm, ArXI 691.8 nm, FeXV 706.2, FeXI 789.4 nm, FeXIII 1074.7 nm, and 1079.8 nm) covering a wide range of temperature (log T$_{eff}$ $\sim$ 5.80 - 6.63). A similar analysis can be applied to the observations of CMEs that will be acquired simultaneously by MLSO/UCoMP and MLSO/K-Cor. In addition, FeX 637.4 nm and FeXI 789.4 nm are temperature-sensitive lines, and FeXIII 1074.7 \& 1079.8 nm are density-sensitive lines. The ratio of these lines, together with the help of atomic spectral line databases (\citealt{1997Dere}), it is possible to infer the 2-D distribution of plasma temperatures and densities inside the CMEs. This study is a testing bed for VELC/Aditya-L1, which will perform
both spectroscopy and imaging of the CMEs in the inner corona in three visible (one continuum centered at 500 nm and two emission lines; FeXIV 530.3 nm \& FeXI 789.2 nm ) and one infrared (FeXIII 1074.7 nm) passbands (\citealt{2021Ritesh}).

\section*{Conflict of Interest Statement}

The authors declare that the research was conducted in the absence of any commercial or financial relationships that could be construed as a potential conflict of interest.

\section*{Author Contributions}

JS led the analysis and carried out the image processing and spectroscopic based investigations. VP and RP planned the analysis and identified the case for analysis. VP and DB assisted in the interpretation of the results. JS prepared the manuscript. All authors took part in the discussion.

 \section*{Funding}
 JS is supported by funds of the Council of Scientific $\&$ Industrial Research (CSIR), India, under file no. 09/0948(12550)/2021-EMR-I. 

\section*{Acknowledgments}
We would like to thank ARIES for providing the computational facilities. Courtesy of the Mauna Loa Solar Observatory, operated by the High Altitude Observatory, as part of the National Center for Atmospheric Research (NCAR). NCAR is supported by the National Science Foundation. The SECCHI data used here were produced by an international consortium of the Naval Research Laboratory (USA), Lockheed Martin Solar and Astrophysics Lab (USA), NASA Goddard Space Flight Center (USA), Rutherford Appleton Laboratory (UK), University of Birmingham (UK), Max-Planck-Institut for Solar System Research (Germany), Centre Spatiale de Li$\grave{e}$ge (Belgium), Institut d'Optique Th$\acute{e}$orique et Appliqu$\acute{e}$e (France), Institut d'Astrophysique Spatiale (France). We also acknowledge SDO team to make AIA data available.


 \section*{Data Availability Statement}
 The MLSO/CoMP, MLSO/K-Cor, SDO/AIA, and STEREO/SECCHI data sets analyzed for this study can be found in their respective data archives under the open data policy. The data sets generated in this study can be made available upon request.

\bibliographystyle{frontiersinSCNS_ENG_HUMS} 
\bibliography{references}



\end{document}